\documentclass[12pt,epsfig,a4]{article} 
\newcommand{\vs}{\vspace{-0.25cm}}
\usepackage{amsmath,graphicx,wasysym}
\usepackage{float}
\begin{document} 
\begin{center}
{\Large{\bf Born series for s-wave scattering length and some exact results} }  

\bigskip

 N. Kaiser \\
\medskip
{\small Physik-Department T39, Technische Universit\"{a}t M\"{u}nchen,
   D-85747 Garching, Germany\\

\smallskip

{\it email: nkaiser@ph.tum.de}}
\end{center}
\medskip
\begin{abstract}
In these notes the Born series for the $s$-wave scattering $a_0$ is calculated for a class of central potentials $V(r)$ up to sixth order in a dimensionless coupling strength $g$. Examples of exponentially decaying potentials as well as truncated potentials involving a single length-scale $a$ are considered. In certain favorable cases the exact result for the $g$-dependent $s$-wave scattering length $a_0=A_0(g) a$ can be given in terms of special functions. The poles of $A_0(g)$ at increasing positive values of $g$ correspond to the thresholds, where $s$-wave bound-states occur successively. A scattering problem, where $A_0(g)$ is solvable in terms of elementary functions, is also presented.

\end{abstract}

\section{Preparation}
Consider the quantum mechanical scattering in a short-range central potential $V(r)$. The time-independent Schr\"odinger equation is equivalently  transformed into an integral equation for the scattering wave-function \cite{qmbook}
\begin{equation}
\psi_k(\vec r\,) = e^{ik z} -{m \over 2\pi \hbar^2} \int d^3r'\, {\exp(i k |\vec r-\vec r\,'|)\over 
|\vec r-\vec r\,'|} V(r') \psi_k(\vec r\,')\,, \end{equation} 
where the first part on the right hand side is the incoming planar wave in $z$-direction, and the second part is the scattered wave. Its asymptotic behavior for $|\vec r\,|\to \infty$  has of the well-known form \cite{qmbook}: complex-valued scattering amplitude $f(k,\theta)$ times an outgoing spherical wave $e^{i kr}/r$. The Born series  is obtained by inserting for  $\psi_k(\vec r\,')$ first the planar wave $e^{i k z'}$, then adding the correction obtained this way linear in $V(r')$, and iterating the procedure. The asymptotic behavior for  $|\vec r\,|\to \infty$ of all these contributions to the scattered wave determines the multiple scattering series \cite{qmbook} for the scattering amplitude  $f(k, \theta)$. In the strict long-wavelength limit  $k\to 0$, where the scattering amplitude $f(k, \theta)$ reduces to the $s$-wave scattering length $a_0=f(0,\theta)$,  this series simplifies drastically, and one obtains the following expansion for $a_0$:
\begin{eqnarray}
 a_0 &=& \Big({-m \over 2\pi \hbar^2}\Big)\!\! \int\!\! d^3r_1 V(r_1) + \Big({-m \over 2\pi \hbar^2}\Big)^2 \!\! \int\!\! d^3r_1 \int \!\!d^3r_2 {V(r_1) V(r_2) \over |\vec r_1-\vec r_2|}\nonumber \\ && + \Big({-m \over 2\pi \hbar^2}\Big)^3 \!\! \int\!\! d^3r_1\!\! \int\!\! d^3r_2 \!\!\int\!\! d^3r_3 {V(r_1) V(r_2)V(r_3) \over |\vec r_1-\vec r_2||\vec r_2-\vec r_3|}+ \dots  \end{eqnarray}
Note that the occurring integrals over a central potential $V(r_i)$, multiplied with the "static propagator" $|\vec r_i-\vec r_j|^{-1}$, are of the same form as electrostatic potentials produced by radially symmetric charge densities. The pertinent angular integral is in each step given by
\begin{equation}
\int d\Omega_i \,{1\over  |\vec r_i-\vec r_j|} = {4\pi \over \text{max}(r_i,r_j)}\,,
\end{equation} 
 where max$(r_i,r_j)$ denotes the maximum of the two radial coordinates $r_i$ and $r_j$. 
 
In order to emphasize the  expansion in a (small) parameter, it is advantageous to factor out the physical constants ($\hbar$ and $m$) and choose for the central potential $V(r)$ the ansatz:
 \begin{equation}V(r) = -{\hbar^2 g \over 2m a^2} f(r/a) \,, \end{equation}
with $a$ the inherent length-scale and  $g$ a dimensionless coupling strength. Moreover, the profile function $f(x)$ specifies the dependence of the central potential $V(r)$ on distance $r$. In this setting the $s$-wave scattering length is necessarily  a multiple of the length-scale $a$, namely $a_0 = A_0 a$. The expansion of the dimensionless quantity $A_0$ in powers of the dimensionless coupling strength $g$ reads
\begin{equation}
A_0(g) = c_1 g + c_2 g^2+c_3 g^3+ c_4 g^4 +c_5 g^5+ c_6 g^6 +\dots\,,
\end{equation} 
where the first few coefficients are computed as 
\begin{eqnarray}
&& c_1 = \int_0^\infty \!\! dx\, x^2f(x)\,, \quad c_2 = \int_0^\infty \!\! dx\, x^2f(x)\Phi(x)\,, \quad c_3 = \int_0^\infty \!\! dx\, x^2f(x) [\Phi(x)]^2\,,  \\ &&c_4 =\int_0^\infty \!\!\!\! dx_1\!  \int_0^\infty \!\!\!\!  dx_2\, {x_1^2 x_2^2\over \text{max}(x_1,x_2)} \Phi(x_1)f(x_1) f(x_2)\Phi(x_2) = 2\! \int_0^\infty \!\! \!\! dx_1\! \int_0^{x_1} \!\!\!\!  dx_2\, x_1 x_2^2 \Phi(x_1)f(x_1) f(x_2)\Phi(x_2) \nonumber .
\end{eqnarray}
Here, $\Phi(x)$ is the "electrostatic potential" produced by the "charge density" $f(x)$:
\begin{equation}
\Phi(x) =\int_0^\infty \!dx' {x'^2\over \text{max}(x,x')}f(x') =\int_0^x \!dx'\, {x'^2\over x}f(x') + \int_x^\infty \!dx'\, x'f(x')\,.
\end{equation}
Higher coefficients $c_j$ involve further factors of $x_i^2 f(x_i)/\text{max}(x_i,x_j)$ in a multiple integral. Note that the increasing powers of the prefactor in the Born series ($-m/2\pi \hbar^2$) and in the potential ($-\hbar^2/2m$), together with $4\pi$ from the angular integrations, have compensated each other at every order.
\section{Exponentially decaying potentials}
In this section the Born series $A_0 =\sum_{j=1}^6 c_j g^j +\dots $ is computed for a variety of central potentials $V(r)$ with an exponential decay.  
\vspace{0.4cm}

\noindent
{\bf A:} The familiar Yukawa potential is physically well motivated in terms of an underlying  one-boson exchange mechanism. The pertinent profile function $f(x)$ and "electrostatic potential" $\Phi(x)$ are in this case
\begin{equation}
f(x) = {e^{-x} \over x}\,, \qquad \Phi(x) = {1 - e^{-x} \over x}\,.
\end{equation}
With this input one derives for the following Born series up to sixth order:
\begin{equation}A_0(g) = g +{g^2 \over 2} + g^3 \ln{4 \over 3} +g^4 \ln{32\over 27} +0.100941276 \,g^5 + 0.06006048 \,g^6+ \dots \end{equation}
Note that in a Feynman diagrammatic picture, where the Yukawa potential stems from a massive one-boson exchange, order $g$ corresponds to tree-level, order $g^2$ to one-loop, order $g^3$ to two-loops,  order $g^4$  to three-loops, and so on. The two-loop coefficient $c_3 = \ln(4/3)= 0.28768207$ has also been found in a momentum-space calculation of twice-iterated one-boson exchange in ref.\cite{mypaper}, while the three-loop coefficient $c_4 = \ln(32/27)=0.16989904$ seems to be a novel result.  It is quite astonishing that even the four-loop coefficient $c_5= 0.1009412761$ can be given in exact form as
\begin{equation}c_5 = {\pi^2 \over 6} +(10+3\ln 2-2\ln 3)\ln2 -{9\over 2}\ln 3 -{5\over 4}\ln 5 +2\text{Li}_2\Big(\!-{1\over 2}\Big) -2\text{Li}_2\Big(\!-{1\over 3}\Big) +2\text{Li}_2\Big(\!-{2\over 3}\Big) \,,\end{equation}
where Li$_2(z) = \sum_{n=1}^\infty z^n/n^2$ denotes the dilogarithmic function.

The series for $A_0(g)$, with its first six terms written in eq.(8), has a finite radius of convergence $(|g|<g_1)$, that is determined as that strength parameter $g_1$, where the threshold for the first $s$-wave bound-state in a Yukawa potential is reached. This particular value $g_1$ can be computed by solving numerically the second-order differential equation for the radial wave-function (multiplied with $x$):
\begin{equation} u''(x) + g {e^{-x}\over x} u(x) = 0\,, \end{equation}
subject to a linear behavior $u(x) \sim x$ at small $x$. Note that $u''(x) + g\, x^{-1} u(x)=0$ has the regular solution in the form of a Bessel-function $u(x) = \sqrt{g\,x } J_1(2\sqrt{g\, x})=g\,x +{\cal O}(x^2)$ \cite{spezfktn}, which also displays a linear behavior. Once the critical strength $g_1$ is reached, the asymptotic behavior of $u(x)$ for large $x$ turns into a  {\it constant}. The convergence of this process is monitored by inspecting, how the first derivative $u'(x)$ approaches zero for large $x$, when fine-tuning more and more digits of the strength parameter $g$. As a result of this numerical exercise, one finds that the first three thresholds $g_1$, $g_2$, $g_3$ for $s$-wave bound-states  in the Yukawa potential lie at the values:
\begin{equation} g_1 = 1.679808\,, \qquad g_2 = 6.44726045\,, \qquad g_3 =14.34202607\,.  \end{equation} 
Note that the earlier determinations $g_1= 1.680,\, g_2 = 6.445,\, g_3 =14.34$ in ref.\cite{jaume} are very close to the present results.   
\vspace{0.4cm}
   
\noindent
{\bf B:} Next, we consider a purely exponentially decreasing potential $V(r) \sim - g \, e^{-r/a}$ with the profile function  
\begin{equation}
f(x) = e^{-x}\,, \qquad \Phi(x) = {2\over x}-\Big(1+{2\over x}\Big)  e^{-x}\,,
\end{equation}
and on this basis, one derives the following Born series up to sixth order
\begin{equation}A_0 = 2\,g +{5\over 4}\,g^2  + {23\over 27}\, g^3 +{677 \over 1152}\,g^4  +{7313\over 18000}\, g^5 + {218491\over  777600}\, g^6+ \dots \end{equation}
For the exponential potential $\sim -g\, e^{-x}$ a little miracle happens, since (at $k=0$) the  second-order linear homogeneous differential equation for the radial wave-function (multiplied with $x$): 
 \begin{equation}
u''(x) + g\, e^{-x} u(x) =0\,,
\end{equation}
possesses analytical solutions in the form of a Bessel-function $u_1(x)=J_0(2 \sqrt{g}\,e^{-x/2})$ and a Neumann-function $u_2(x)=N_0(2 \sqrt{g}\,e^{-x/2})$ \cite{spezfktn} of index $0$. The proper linear combination is of these two is found by demanding $u(0)=0$ and an asymptotic behavior $u(x) \simeq x+A_0$ for $x\to \infty$. The latter asymptotic behavior $u(x) \simeq x(1+A_0/x)$  is provided by the logarithmic singularity of the Neumann-function $N_0(z)=2[\gamma_E +\ln(z/2)]/\pi +{\cal O}(z^2\ln z) $ near $z=0$. Putting the pieces together, the  resulting exact expression for the $s$-wave scattering length in the exponential  potential reads: 
\begin{eqnarray}
&& A_0(g) = {\pi N_0(2 \sqrt{g}) \over J_0(2 \sqrt{g})} -\ln g -2 \gamma_E\,, \quad \text{for} \,\,\, g>0\,, \nonumber \\   && A_0(g) =- {2K_0(2 \sqrt{-g}) \over I_0(2 \sqrt{-g})} -\ln(- g) -2 \gamma_E\,, \quad \text{for} \,\,\, g<0\,,
\end{eqnarray}\begin{figure}[h]
	\centering
		\includegraphics[scale=1.2,clip]{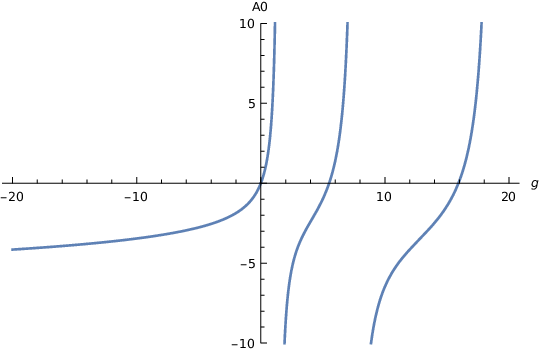}
\caption{Scattering length $A_0=a_0/a$ versus $g$ for exponential potential $V(r)\sim - g\, e^{-r/a}$.}
\end{figure}\noindent with $\gamma_E = 0.577215665$ the Euler-Mascheroni number. 
In the second formula (for $g<0$), $K_0$ and $I_0$ are modified Bessel-functions \cite{spezfktn} that arise from the continuation of $N_0$ and $J_0$ to imaginary argument. The behavior of  $A_0(g)$ is shown in Fig.\,1 for both positive and negative values of the strength parameter $-20<g<20$. One observes poles at positive values of $g$, which mark the thresholds for successive $s$-wave bound-states in the attractive exponential potential. According to exact formula in eq.(16), these thresholds can be  identified with the zeros $z_j$ of the Bessel-function $J_0(z)$ through the relation  $g_j= z_j^2/4$. The first six threshold values read
\begin{equation}
g_1\!=\! 1.445796\,,\quad  g_2\!=\! 7.617816\,, \quad g_3 \!=\! 18.72175\,, \quad g_4 \!=\! 34.76007\,, \quad g_5\!=\! 55.73308\,,\quad  g_6 \!=\!81.64084\,. \end{equation}
Note that the numerical method, employed before to construct $g_1, g_2, g_3$ for the Yukawa potential, reproduces $g_1= 1.4457965$, $g_2= 7.6178156$, $g_3 = 18.7217517$  for the profile $f(x) =e^{-x}$. 
\vspace{0.4cm}

\noindent
{\bf C:} As a further example, we choose an exponentially decreasing potential with a soft core:
\begin{equation}
	f(x) = x\, e^{-x}  \,, \qquad \Phi(x) = {6\over x}-\Big( {6\over x}+4+x\Big) e^{-x}\,,
\end{equation}
and derive the following Born series up to sixth order

\begin{equation}
	A_0(g) = 6\,g +{33\over 4}\,g^2  + {5873\over 486}\, g^3 +17.835277 g^4+ 26.351768\, g^5  + 38.941427\,g^6  + \dots \,,
\end{equation}
where all expansion coefficients are positive rational numbers. The radius of convergence $|g|<g_1$ of the series for $A_0(g)$ written in eq.(19) is determined by searching the threshold for the first $s$-wave bound-state in the potential $V(r) \sim - g\, x \,e^{-x}$. One finds with high numerical accuracy the values
\begin{equation} g_1 = 0.676681997\,, \qquad g_2 =4.30765405\,, \qquad g_3 = 21.020136626\,. \end{equation}
\vspace{0.cm}

\noindent 
{\bf D:} Next, we choose an exponentially decreasing potential with a very soft core, taking the profile function
\begin{equation}
	f(x) = x^2 e^{-x}  \,, \qquad \Phi(x) = {24\over x}-\Big( {24\over x}+18+6x+x^2\Big) e^{-x}\,,
\end{equation}
and derive the following Born series up to sixth order
\begin{equation}
	A_0(g) = 24\,g +{837\over 8}\,g^2  + 479.53818\, g^3 +2209.6103\, g^4+10188.272 \, g^5  + 46981.234\,g^6  + \dots \,,
\end{equation}
where all expansion coefficients are (large) rational numbers. The radius of convergence $|g|<g_1$ of the series for $A_0(g)$ in eq.(22) is again determined by searching the threshold for the first $s$-wave bound-state in the potential $V(r) \sim -g\, x^2 e^{-x}$. One finds with high numerical accuracy the values
\begin{equation} g_1 = 0.216855259 \,, \qquad g_2 = 1.580333075\,, \qquad g_3= 8.027999134\,. \end{equation}
\vspace{0.cm}

\noindent
{\bf E:} Finally, one can consider an exponentially decreasing potential with a more singular behavior at small distances:
\begin{equation}
	f(x) = {e^{-x}\over x^2}  \,, \qquad \Phi(x) = {1-e^{-x}\over x}+\Gamma(0,x)\,,
\end{equation}
where $\Gamma(0,x) = \int_x^\infty \!d\xi\, \xi^{-1} e^{-\xi} = -\ln x -\gamma_E +x -x^2/4 +{\cal O}(x^3)$ denotes the incomplete $\Gamma$-function \cite{spezfktn}.  With this input one derives  the following Born series up to sixth order:
\begin{equation}
	A_0(g) = g +g^2 \ln 4  + 2.98036548\, g^3 +7.68193343\, g^4+21.86973
 \, g^5  + 66.2866\,g^6  + \dots
\end{equation}
The profile $f(x)= e^{-x}/x^2$ is too singular near the origin and therefore the coupling strength $g$ must be restricted to negative or small positive values $g< g_\text{crit}=1/4$. This feature concerning an instability  is further elaborated on in the next section, when analyzing the truncated $1/r^2$-potential.

\section{Truncated potentials}
In this section we analyze short-range potentials, that vanish outside a sphere of radius $a$.
\vspace{0.4cm}
 
\noindent 
{\bf A:}  A well-known textbook problem \cite{qmbook} is the quantum mechanical scattering off a spherical potential-well, represented by the profile  
\begin{equation}
f(x) = \theta(1-x) \,, \qquad \Phi(x) = {1\over 2}-{x^2\over 6}\,,
\end{equation}
and now the knowledge of the "electrostatic potential" $\Phi(x)$ is needed only in the inside region $0<x<1$. With little effort, one obtains the following Born series for the $s$-wave scattering length in a spherical potential-well up to sixth order:
\begin{equation}
A_0(g) = {g \over 3} +{2\over 15}\,g^2  + {17\over 315}\, g^3 +{62 \over 2835} g^4+ {1382 \over 155925}\, g^5  + {21844 \over 6081075}\,g^6  + \dots
\end{equation}
The analytical solution to the differential equation for the radial wave-function $u''(x)+ g\, u(x)=0$ (in the region $0<x<1$) subject to the boundary condition $u(0)=0$ is either a sine-function, $\sin(\sqrt{g}x)$,  or a sine-hyperbolic function, $\sinh(\sqrt{-g}x)$.  After matching the logarithmic derivative  of $u(x)$ at $x=1$ to that of the linear form $u(x)=x+A_0$ outside, one obtains the generic formula $A_0 = u(1)/u'(1)-1$ for the $s$-wave scattering length in any truncated potential.  In the present case one reproduces the well-known result   
\begin{eqnarray}
&& A_0(g) = -1+ {1\over \sqrt{g}} \tan \sqrt{g} \,, \quad \text{for} \,\,\, g>0\,, \nonumber \\   && A_0(g) =-1 + {1\over \sqrt{-g}}  \tanh \sqrt{-g}\,, \quad \text{for} \,\,\, g<0\,,
\end{eqnarray}
and, as a nice check, the Taylor expansion of the tangent-function confirms the series in eq.(27). The analytical result for $A_0(g)$ written in eq.(28) is shown in Fig.\,2 in the region $-40<g<62$.  
The poles of $A_0(g)$  at the  values $g_j = \pi^2(j-1/2)^2$ (where $\tan\sqrt{g_j}$ diverges) correspond to the well-known thresholds for the occurrence of successive $s$-wave bound-states in a spherical potential-well. The first six values of this infinite sequence read 
\begin{equation}
g_1 \!=\! 2.467401\,,\quad  g_2\!=\! 22.20661\,, \quad g_3 \!=\!  61.68503\,, \quad g_4 \!=\! 120.9027\,, \quad g_5\!=\! 199.8595\,, \quad g_6 \!=\! 298.5555\,.
\end{equation}

\begin{figure}[h]
	\centering
		\includegraphics[scale=1.3,clip]{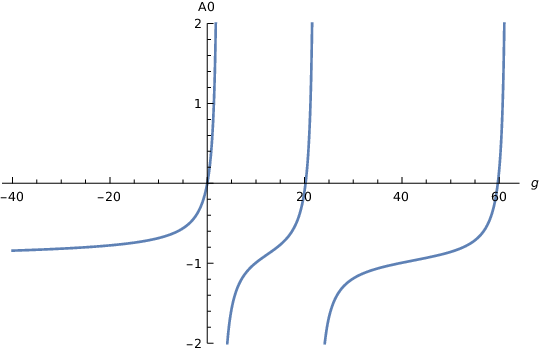}
		\caption{$A_0$ as a function of $g$ for spherical well  potential.}
\end{figure}

\noindent
{\bf B:}
It is also interesting to study  the truncated $1/r^2$-potential with the profile function
\begin{equation}
f(x) = \Big( {1\over x^2}-1\Big) \theta(1-x) \,, \qquad \Phi(x) = {1\over 2} +{x^2\over 6} -\ln x\,,
\end{equation}
and this input leads to the following Born series  for the $s$-wave scattering length up to sixth order
\begin{equation}
A_0(g) = {2 \over 3}\,g +{56\over 45}\,g^2  + {14312\over 4725}\, g^3 +8.364150\, g^4+ 24.881527\, g^5  + 77.758378\,g^6  + \dots 
\end{equation}
The differential equation for the radial wave-function in the inner region $(0<x<1)$:  
\begin{equation}
u''(x) + g \Big({1\over x^2}-1\Big) u(x) = 0\,,
\end{equation}
possesses a regular solution in the form of a Bessel-function $u(x) = \sqrt{x} J_\nu(\sqrt{-g} x)$ with index $\nu = \sqrt{1/4-g}$. Note that the strength parameter $g$ has to be constrained to $g\leq 1/4$ in order to evade the instability of the quantum mechanical  problem with a too attractive $1/r^2$-potential. For comparison the regular solution of $u''(x)+g\, x^{-2}u(x)=0$ is for all $g\le 1/4$ given by $u(x) = x^{\nu +1/2}$. Under this restriction  on $g$ the logarithmic derivative $u'(x)/u(x)$ at $x=1$ determines the dimensionless $s$-wave scattering length as $A_0= u(1)/u'(1)-1$. In the present case this leads to the expression: 
\begin{equation}
A_0(g) = \bigg\{  {1+\sqrt{1-4g}\over 2} -\sqrt{-g}\,  {J_{\nu+1}(\sqrt{-g}) \over J_\nu(\sqrt{-g})}\bigg\}^{-1}-1\,, \qquad \nu =  {\sqrt{1-4g}\over 2}\,,
\end{equation} 
where the second term in curly brackets, $-\sqrt{-g} J_{\nu+1}(\sqrt{-g})/J_\nu( \sqrt{-g} )$, means  $\sqrt{g} I_{\nu+1}(\sqrt{g})/I_\nu(\sqrt{g})$ for $0<g<1/4$.  The exact result for $A_0(g)$ written in eq.(33) is shown in Fig.\,3 for the parameter range $-5<g <1/4$. As already visible in Fig.\,2, and valid for any truncated potential, the scattering length approaches with increasing repulsion the hard-core value: $\lim_{g\to - \infty} A_0(g) = -1$. On the positive side  the boundary value, before instability sets in, is given by
\begin{equation}
A_0(1/4) = {I_0(1/2) - I_1(1/2) \over I_0(1/2) + I_1(1/2)} = 0.6095845  \,.
\end{equation}

\begin{figure}[h]
\centering
	\includegraphics[scale=1.3,clip]{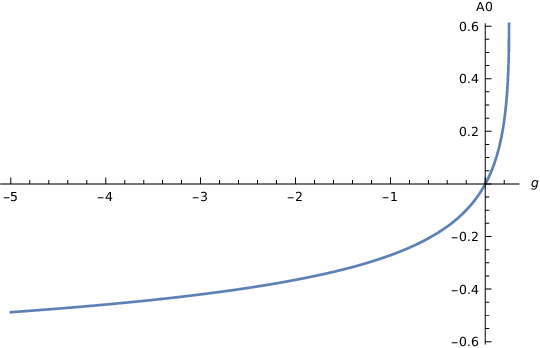}
\caption{$A_0$ as a function of $g$ for truncated $1/r^2$-potential, restricted to range $g\leq 1/4$.}
\end{figure}

\noindent
{\bf C:}
Next, we consider the truncated Coulomb-potential with profile function
\begin{equation}
	f(x) = \Big( {1\over x}-1\Big) \theta(1-x) \,, \qquad \Phi(x) = {1-x\over 2 }+{x^2\over 6} \,,
\end{equation} 
and derive the following Born series up to sixth order
\begin{equation}
A_0(g) = {g \over 6} +{g^2\over 20}  + {241\over 15120}\, g^3 +{233 \over 45360}\, g^4+ {g^5\over 602.7393 }  + {g^6 \over 1865.2312}  + \dots 
\end{equation}
\begin{figure}[H]
\centering
	\includegraphics[scale=1.3,clip]{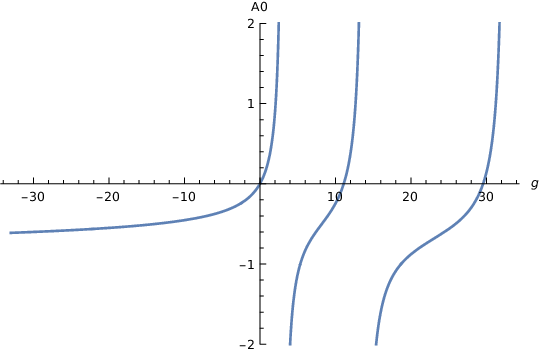}
\caption{$A_0$ as a function of $g$ for truncated Coulomb-potential.}
\end{figure}

\noindent 
Now, the differential equation  $u''(x)+g(x^{-1}-1)u(x) =0$ for the radial wave-function possesses  a regular solution of the form $u(x) =  x\, e^{-\sqrt{g}\,x}\, _1F_1(1-\sqrt{g}/2,2;2\sqrt{g}\,x)$, including a confluent hypergeometric series \cite{spezfktn}. Based on this solution the exact expression for the $s$-wave scattering length in the truncated Coulomb-potential $A_0(g)= u(1)/u'(1)-1$ takes the form:
\begin{equation}
A_0(g) = \bigg\{1 - \sqrt{g} + {(\sqrt{g} - g/2)\,\, _1F_1(2 - \sqrt{g}/2, 3; 2 \sqrt{g}) \over 
 _1F_1(1 - \sqrt{g}/2, 2; 2 \sqrt{g})}\bigg\}^{-1}-1\,. 
\end{equation}
By solving the equation, that the numerator of the $g$-dependent function in curly brackets is zero, one determines the threshold for successive $s$-wave bound-states in the truncated Coulomb-potential. The first  six zeros are computed as: 
\begin{equation}
g_1 \!=\! 3.094169\,,\quad g_2 \!=\! 13.88101\,, \quad g_3 \!=\! 32.62548\,, \quad
g_4\!=\!  59.35213\,, \quad g_5 \!=\! 94.068586\,, \quad g_6 \!=\! 136.77831\,. 
\end{equation}
The expression for $A_0(g)$ written in eq.(37) is plotted in Fig.\,4 in the range $-33<g<33$. 
\vspace{0.4cm}

\noindent
{\bf D:} Let us form the difference of the truncated $1/r^2$-potential and the truncated  Coulomb-potential by working with the profile function
\begin{equation}
	f(x) = \Big( {1\over x^2}-{1\over x}\Big) \theta(1-x) \,, \qquad \Phi(x) = {x\over 2 }- \ln x\,.
\end{equation}
With this input one derives the following Born series up to sixth order
\begin{equation}
	A_0(g) = {g \over 2} +{5\over 6}\,g^2  + {275\over 144}\, g^3 +{21949 \over 4320}\, g^4+ 17.73814\,g^5\  + 45.22973\,g^6  + \dots \,,
\end{equation}
where all expansion coefficients are positive rational numbers. The pertinent differential equation $u''(x)+g(x^{-2}-x^{-1}) u(x)=0$ has the regular solution $u(x)= \sqrt{x} J_\mu(2 \sqrt{-g\, x})$ involving a Bessel-function of index $\mu = \sqrt{1-4g}$.  On this basis one obtains the following exact result for the $s$-wave scattering length
\begin{equation}
	A_0(g) = \bigg\{  {1+\sqrt{1-4g}\over 2} -\sqrt{-g}\,  {J_{\mu+1}(2\sqrt{-g}) \over J_\mu(2\sqrt{-g})}\bigg\}^{-1}-1\,, \qquad \mu =  \sqrt{1-4g}\,,
\end{equation} 
\begin{figure}[H]
\centering
\includegraphics[scale=1.3,clip]{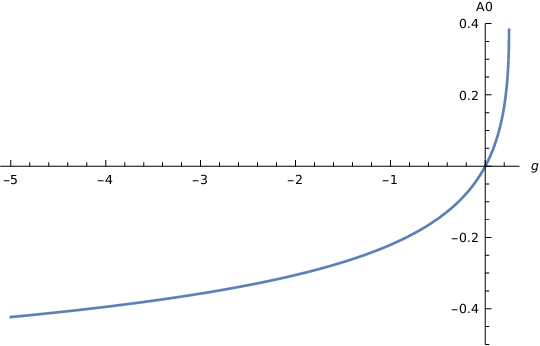}
\caption{$A_0$ as a function of $g$ for difference of truncated  $1/r^2$- and Coulomb-potential.}
\end{figure}

\noindent
where the second term in curly brackets, $-\sqrt{-g} J_{\mu+1}(2\sqrt{-g})/J_\mu( 2\sqrt{-g} )$, means  $\sqrt{g} I_{\mu+1}(2\sqrt{g})/I_\mu(2\sqrt{g})$ for $0<g<1/4$. The expression for $A_0(g)$ written in eq.(41) is shown in Fig.\,5 over the parameter range $-5<g <1/4$. The boundary value, where the slope diverges and  instability sets in, is given by 
\begin{equation}
	A_0(1/4) = {I_0(1) - I_1(1) \over I_0(1) + I_1(1)} = 0.38275296  \,.
\end{equation}
 
\noindent
{\bf E:} A further interesting example is the truncated oscillator potential with the profile function
\begin{equation}
	f(x) = (1-x^2) \theta(1-x) \,, \qquad \Phi(x) = {1\over 4}-{x^2\over 6 }+{x^4\over 20} \,,
\end{equation}
and one derives the following Born series up to sixth order
\begin{equation}
A_0(g) = {2 \over 15}\, g +{8\over 315}\,g^2  + {1432\over 289575}\, g^3 +{g^4 \over 1036.0372} + {g^5\over 5306.4658 }  + {g^6 \over 27178.112}  + \dots 
\end{equation}
The solution of the differential equation $u''(x)+g(1-x^2)u(x) =0$ can be found in terms of parabolic cylinder functions $D_\nu(\sqrt{\pm 2} g^{1/4}x)$ \cite{spezfktn} with respective indices $\nu = (-1\pm \sqrt{g})/2$. From the pertinent linear combination which satisfies $u(0)=0$, one gets through the formula $A_0(g) = u(1)/u'(1)-1$ a very lengthy expression that will not be presented here. The pertinent numerical result is shown in Fig.\,6 for the parameter range $-110<g<110$. The first six poles of $A_0(g)$ corresponding to thresholds for $s$-wave bound-states in the truncated oscillator potential lie at: 
\begin{equation} 
g_1 \!=\! 5.121669\,, \quad g_2 \!=\! 39.66084\,, \quad g_3\!=\! 106.2492\,, \quad g_4\!=\!  204.8561\,, \quad g_5 \!=\! 335.4732\,, \quad g_6\!=\!  498.0971\,.
\end{equation}
\begin{figure}[h]
\centering
\includegraphics[scale=1.3,clip]{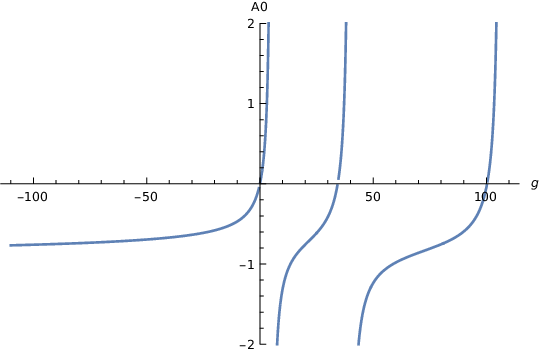}
\caption{$A_0$ as a function of $g$ for truncated oscillator potential.}
\end{figure}

\noindent
{\bf F:} As a further example we consider the truncated linear potential with the profile function
\begin{equation}
f(x) = (1-x) \theta(1-x) \,, \qquad \Phi(x) = {1-x^2\over 6 }+{x^3\over 12} \,,
\end{equation}
and derive the following Born series  up to sixth order:
\begin{equation}
A_0(g) = {g \over 12} +{13\over 1260}\,g^2  + {17\over 12960}\, g^3 +{g^4 \over 5977.6798} + {g^5\over 46852.232 }  + {g^6 \over 367200.57}  + \dots 
\end{equation}
The solution of the differential equation $u''(x)+g(1-x)u(x) =0$ for the radial wave-function (in the region $0<x<1$) subject to the boundary condition $u(0)=0$ is given by a linear combination of Airy-functions \cite{spezfktn}.  Its logarithmic derivative at $x=1$ leads to the following exact expression for the $s$-wave scattering length in a truncated linear potential:
\begin{equation}
A_0(g) = {\big[ \sqrt{3}\text{Ai}(-g^{1/3})- \text{Bi}(-g^{1/3})\big] \Gamma(1/3) \over (3g)^{1/3} \big[ \sqrt{3}\text{Ai}(-g^{1/3})+\text{Bi}(-g^{1/3})\big] \Gamma(2/3)}-1 \,,
\end{equation}
which allows to confirm the Taylor series expansion written in eq.(47). The numerical result for $A_0(g)$ written in eq.(48) is shown in Fig.\,7 over the large parameter range $-150<g<150$. The  poles of $A_0(g)$ at positive $g$ demarcate the thresholds for successive $s$-wave bound-states in the truncated linear potential. The first six zeros of the pertinent equation, $\sqrt{3}\text{Ai}(-g^{1/3})+\text{Bi}(-g^{1/3})=0$, are  computed as
\begin{equation} g_1 \!=\! 7.837347\,, \quad g_2 \!=\! 55.97703\,, \quad g_3\!=\!148.5083\,, \quad g_4\!=\!285.4502\,, \quad g_5\!=\!    466.8047\,,\quad g_6 \!=\!692.5722\,.
\end{equation}

\begin{figure}[h]
\centering
\includegraphics[scale=1.3,clip]{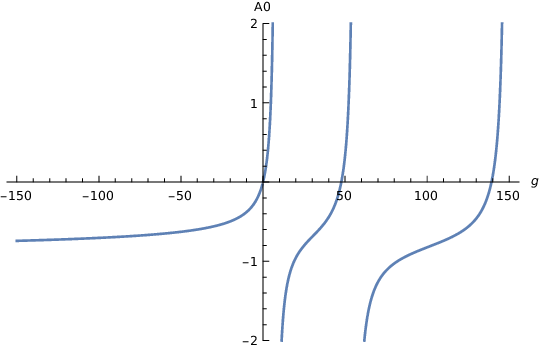}
\caption{$A_0$ as a function of $g$ for truncated linear potential.}
\end{figure}

\noindent 
{\bf G:} As a last example we consider a truncated  potential of cosinus shape with the profile function
\begin{equation}
f(x) = {\pi^2 \over 4}\cos{\pi x\over 2}\,  \theta(1-x) \,, \qquad \Phi(x) = {\pi \over 2} + \cos{\pi x\over 2}-{4 \over \pi x}\sin{\pi x\over 2}\,,
\end{equation}
where the prefactor $\pi^2/4$ as been included for convenience. One derives the following Born series  up to sixth order:
\begin{equation}
A_0(g) = g\,{\pi^2-8 \over 2\pi} +g^2\, {7\pi^2-66\over 24}+g^3\Big({\pi^3 \over 6}-{17\pi\over 12}-{56\over 27\pi}\Big) +{g^4 \over 39.594202} + {g^5\over 89.212563 }  + {g^6 \over 201.00296}  + \dots 
\end{equation}
The differential equation $u''(x)+g f(x)u(x)=0$ for the radial wave-function is now solved by the odd Mathieu-function \cite{spezfktn} with characteristic value $0$  and parameter $-2g$ of the argument $\pi x/4$. The corresponding exact result for the $s$-wave scattering length $A_0(g) = u(1)/u'(1)-1$ is shown in Fig.\,8 over the parameter range $-90<g<90$. The first six poles, belonging to the thresholds for $s$-wave bound-states in the truncated cosinus-shaped potential, lie at values of the coupling strength: 
\begin{equation} g_1= 2.253065 \,, \quad g_2=17.14564 \,, \quad g_3=45.79166 \,, \quad g_4 = 88.18843\,, \quad g_5 =144.3357 \,, \quad g_6 = 214.1899\,.
\end{equation}
\begin{figure}[h]
\centering
\includegraphics[scale=1.3,clip]{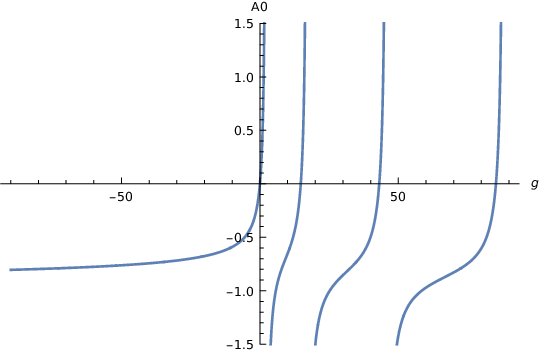}
\caption{$A_0$ as a function of $g$ for truncated potential of cosinus shape.}
\end{figure}
\section{Further  solvable potential scattering problems} 
\noindent 
{\bf A:} Here, we present and discuss another solvable scattering problem, where the potential has the shape of a squared Lorentz-curve, i.e. the profile function is
\begin{equation} f(x) = {1\over (1+x^2)^2 } \,, \qquad \Phi(x) = {\arctan x\over 2x }\,.
\end{equation}
The pertinent second-order differential equation for the radial wave-function 
\begin{equation} u''(x) + {g\over (1+x^2)^2 } u(x) = 0 \,,
\end{equation}
possesses the analytical solution $u(x) = \sqrt{1+x^2}\sin(\sqrt{1+g}\arctan x)$, subject to the boundary condition $u(0)=0$. From its asymptotic behavior $u(x) \simeq const(x+A_0)$ for $x \to \infty$, one can deduce with little effort the following exact result for the dimensionless $s$-wave scattering length:
\begin{eqnarray}
&& A_0(g) = -\sqrt{1+g} \cot\Big( {\pi \over 2 } \sqrt{1+g}\Big) \,, \quad \text{for} \,\,\, g>-1\,, \nonumber \\   && A_0(g) =-\sqrt{-1-g} \coth\Big( {\pi \over 2 } \sqrt{-1-g}\Big) \,, \quad \text{for} \,\,\, g<-1\,,
\end{eqnarray}
with the intermediate value $A_0(-1) = -2/\pi= -0.63661977$. The behavior of $A_0(g)$ written in eq.(55) is shown in Fig.\,9 over  the wide parameter range $-100<g<100$. The poles of $A_0(g)$, which demarcate the thresholds for $s$-wave bound-states in a potential of squared Lorentz shape, lie at coupling strengths $g_j= 4j^2-1$. This sequence of integer values reads up to $j=6$:
\begin{equation} g_1=3\,, \qquad g_2 = 15\,, \qquad g_3 =35\,, \qquad g_4 = 63\,, \qquad g_5 = 99\,, \qquad  g_6 = 143\,.\end{equation}

\begin{figure}[h]
\centering
	\includegraphics[scale=1.3,clip]{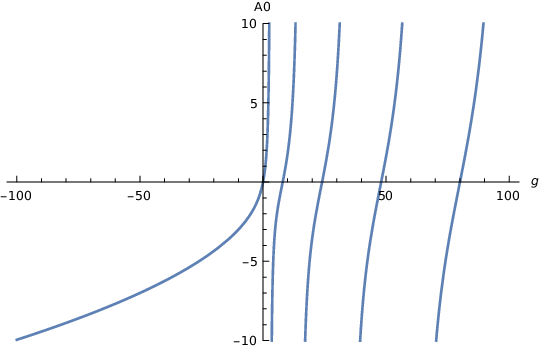}
\caption{$A_0$ as a function of $g$ for potential of squared Lorentz shape.}
\end{figure}
The Taylor expansion of $A_0(g)$ written in eq.(55) gives  the following Born series up to sixth order
\begin{eqnarray} A_0(g) &\!\!=\!\!& g\,{\pi \over 4} + g^2\, {\pi \over 16}+ g^3\, {\pi(\pi^2 -6)\over 192} +g^4\,{\pi(15-\pi^2)\over 768}  +g^5\,{\pi(2\pi^4+5\pi^2-210)\over 15360} + g^6\, {\pi (105-\pi^4)\over 10240}\dots\nonumber \\
	&\!\!=\!\!&  0.7853982\, g + 0.1963495\, g^2 + {g^3\over 15.793733} + { g^4\over 47.649735 }  + {g^5 \over 143.10164}+ {g^6\over 429.39433} + \dots 
\end{eqnarray}
It is a good check to reproduce the occurring expansion coefficients $c_1, \dots , c_6$ by evaluating the multiple integrals in eq.(6) with the profile functions $f(x)$ and $\Phi(x)$ written in eq.(53).
\vspace{0.4cm}

\noindent{\bf B:} Reconsider the $1/r^2$-potential, but now  with a discontinuity at the boundary, by taking the profile function
\begin{equation} f(x) = {\theta(1-x) \over x^2}\,, \qquad \Phi(x) =1 - \ln x\,. \end{equation}
The regular solution of $u''(x)+g\, x^{-2} u(x)=0$ is $u(x) = x^{(1+\sqrt{1-4g})/2}$ for restricted strength parameter $g\leq1/4$, and this leads via $A_0=u(1)/u'(1)-1$ to the following result for the $s$-wave scattering length
\begin{equation} A_0(g) = { 1 - \sqrt{1-4g} \over 1 + \sqrt{1-4g}} \,, \qquad \text{for}\qquad g\leq {1\over 4}\,. \end{equation}
The corresponding Born series reads up to tenth order 
\begin{equation} A_0(g) = g+2\, g^2+5\, g^3+14\, g^4 +42\, g^5+132\, g^6+429\, g^7 + 1430\, g^8 +4862\, g^9+ 16796 \,g^{10}+ \dots \ \end{equation}
The expression for $A_0(g)$ written in eq.(59) is shown in Fig.\,10 over the parameter range $-6<g<1/4$. 
\begin{figure}[h]
	\centering
	\includegraphics[scale=1.3,clip]{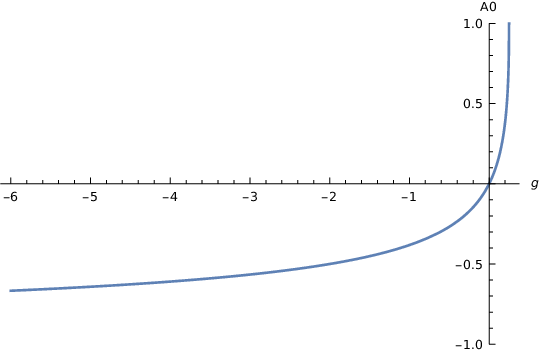}
	\caption{$A_0$ as a function of $g$ for $1/r^2$-potential with discontinuity at $r=a$.}
\end{figure}

\noindent
{\bf C:} In order that a truncated potential with a power-like drop-off $\sim r^{-\alpha}$ can support bound-states, the exponent $\alpha$ must be smaller than $2$. For this reason we consider the $n$-dependent profile function
\begin{equation} f(x) = {\theta(1-x) \over x^{2-1/n}}\,, \qquad \Phi(x) =n -{n^2 \over n+1}\, x^{1/n} \,. \end{equation}  
\begin{figure}[H]
	\centering
	\includegraphics[scale=1.3,clip]{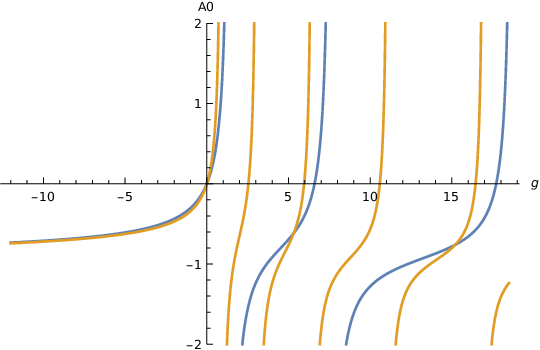}
	\caption{$A_0$ as a function of $g$ for profiles $f(x)=x^{-1}$ (dark) and $f(x) = x^{-3/2}$ (light).}
\end{figure}
\noindent
The regular solution of $u''(x)+g\, x^{1/n-2} \,u(x)=0$ is given by a Bessel-function $u(x) = \sqrt{x} J_n\big(2n \sqrt{g\, x^{1/n}}\big)$, and this leads via $A_0=u(1)/u'(1)-1$ to the following exact result for the $s$-wave scattering length
\begin{eqnarray} && A_0(g) = { J_n(2n \sqrt{g}) \over \sqrt{g}\, J_{n-1}(2n \sqrt{g}) } -1 \,, \qquad \text{for}\qquad g>0\,, \nonumber \\ && A_0(g) = { I_n(2n \sqrt{-g}) \over \sqrt{-g}\, I_{n-1}(2n \sqrt{-g}) } -1 \,, \qquad \text{for}\qquad g<0\,,\end{eqnarray}
where $I_n(z)$ are modified Bessel-functions obtained from $J_n(z)$ by extension to  imaginary argument. The curves in Figs.\,11 and 12 display these results for $n=1,2$ and $n=3,4$, respectively. \begin{figure}[H]
	\centering
	\includegraphics[scale=1.3,clip]{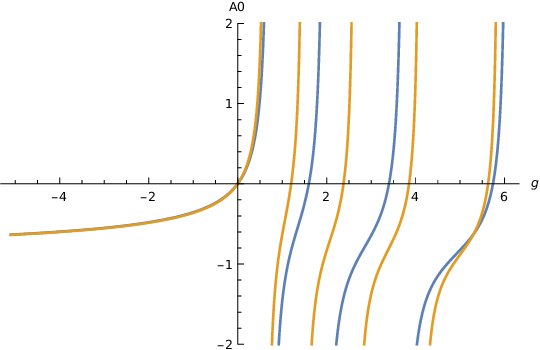}
	\caption{$A_0$ as a function of $g$ for profiles $f(x)=x^{-5/3}$ (dark) and $f(x) = x^{-7/4}$ (light).}
\end{figure}
\noindent
The Taylor expansion around $g=0$ gives the following Born series up to sixth order
\begin{eqnarray} A_0(g) &=& g\, {n \over n\!+\!1} + g^2{2n^2 \over (n\!+\!1) (n\!+\!2)}+ g^3{n^3(5n\!+\!6) \over (n\!+\!1)^2 (n\!+\!2)(n\!+\!3)}
+ g^4 {2n^4(7n\!+\!12) \over (n\!+\!1)^2 (n\!+\!2)(n\!+\!3)(n\!+\!4)} \nonumber \\ && + g^5{ 2n^5(21n^3\!+\!118n^2\!+\!214n\!+\!120)\over(n\!+\!1)^3 (n\!+\!2)^2 (n\!+\!3)(n\!+\!4) (n\!+\!5)} + g^6{ 4n^6 (33n^3\!+\!230n^2\!+\!552n\!+\!360)\over
(n\!+\!1)^3 (n\!+\!2)^2(n\!+\!3)(n\!+\!4) (n\!+\!5)  (n\!+\!6)}+ \dots \nonumber \\  \end{eqnarray} 
The poles of $A_0(g)$, which demarcate the thresholds for successive $s$-wave bound-states in a truncated potential with $r^{1/n-2}$ drop-off, occur at  values of the coupling strength: $g_j^{(n)} = \big(z_j^{(n-1)}/2n\big)^2$, where $z_j^{(n-1)}$ denotes the $j$-th non-trivial zero of the Bessel-function $J_{n-1}(z)$. Interestingly, for $n=1$ these are the same values as found for the exponentially decreasing potential $f(x)=e^{-x}$ in sect.\,2, eq.(17).
\vspace{0.4cm}

This concludes our presentation of results for the $s$-wave scattering length $A_0(g)$ for selected classes of short-range central potentials $V(r)\sim - g\, f(x)$. Of course, one could continue these analytical and numerical calculations for countless other profiles functions $f(x)$. A closed-form expression for $A_0(g)$ can be expected to exist only in rare cases.

\end{document}